# Evaporation of sessile water droplets on horizontal and vertical bi-phobic patterned surfaces

Wenliang Qi,[†,‡] Junhui Li,[†] Patricia B. Weisensee[†,§,*]

[†] Department of Mechanical Engineering & Materials Science, Washington University in St. Louis, MO, USA
[‡] College of Power and Energy Engineering, Harbin Engineering University, Harbin, China
[§] Institute of Materials Science and Engineering, Washington University in St. Louis, MO, USA

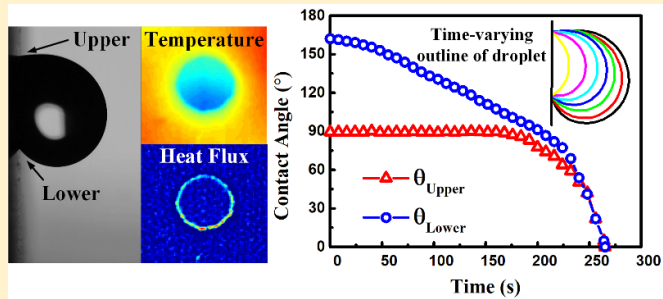

**ABSTACT:** This paper presents an experimental study on thermal transport to single water droplets evaporating on heated bi-phobic surfaces consisting of a superhydrophobic matrix with a circular hydrophobic pattern with strong contact line pinning. A single water droplet of 8 μl volume is placed on a preheated surface and allowed to evaporate in an open laboratory environment. We investigate the influence of substrate orientation (horizontal and vertical) on evaporation dynamics. Using optical and infrared imaging, we report droplet fluid dynamics and heat transfer characteristics of the evaporating droplet. Overall, evaporation is more efficient on the vertical surface, exhibiting higher total heat transfer rates and up to 10% shorter evaporation times. Counterintuitively, on the vertical surface, the substrate-droplet interfacial heat flux was higher near the lower contact line than in the upper region, despite a high contact angle and an expected wedge effect at the bottom. At the same time, the temperature is colder in the lower part of the droplet. We attribute this apparent anomaly to the competition between sensible heating and evaporation, and a modified convective flow signature (both within the droplet and the gas phase) compared to a horizontal surface. We also show that the thermal signature becomes uniform once the contact angles at the upper and lower contact lines become equal towards the end of the evaporation process. Insights from this work can guide the design of spray cooling devices, or be used to alter particle deposition patterns during evaporation-based fabrication techniques and ink-jet printing.

## ■ INTRODUCTION

The evaporation of droplets has shown great potential in a myriad of industrial applications, including spray coating,[1,2] spray cooling,[3,4] inkjet printing,[5,6] bio-sensing[7] and thermal management of electronic devices.[8–10] Droplet evaporation is affected by many parameters, such as substrate thermal properties,[11–13] roughness,[14,15] and wettability.[16–18] Many studies have explored the interaction between droplet evaporation dynamics and the wetting characteristics of the surfaces.[16,19] For example, Pan et al.[18] found that surface wettability can influence the evaporation flux along the interface of a droplet, resulting in different evaporation rates as a function of contact angle. The majority of these studies, however, considered the evaporation on single-wettability surfaces.[14,20–28] For example, in another study, Pan et al.[21] investigated the conjugate heat transfer mechanism at the contact line of non-wetting droplets and found that the maximum evaporation flux occurs in the three-phase contact line region. Gibbons et al.[25] provided the spatial distribution of the substrate-droplet interfacial heat flux as well as the time-varying heat flux distribution for a full evaporation event on a superhydrophobic substrate.

Wettability-patterned surfaces usually contain co-located areas of two or more different wettabilities (superhydrophobic, hydrophobic, hydrophilic, or superhydrophilic), and offer the opportunity to tailor droplet dynamics to achieve the desired wettability and functionality of a surface.[29–33] Based upon the different evaporation regimes on (super)hydrophilic surfaces, on which droplets evaporate with nearly constant contact radius (CCR), and (super)hydrophobic surfaces, which exhibit predominantly a constant contact angle (CCA) mode, differences in evaporation dynamics on wettability-patterned surfaces have been observed in a few previous studies.[34,35] For example, Jansen et al.[35] found that elongated droplets on chemically stripe-patterned surfaces evaporate faster than more spherical droplets due to an increased contact line length.

Limited studies have investigated the evaporation dynamics for a deformed droplet due to gravitational effects on horizontal or tilted substrates.[36–40] In mathematical analyses (analytical or numerical), the non-symmetric droplet shape due to gravitational deformation increases the complexity of the modeling process significantly. Zubkov et al.[41] developed a model for heating and evaporation processes of a liquid spheroidal (prolate and oblate) droplet. Du and Deegan[37] provided a comprehensive analysis on solute deposition by droplet evaporation of droplets with different initial volumes and substrate inclinations, and found that the deposit can be larger at either the upper or lower contact line depending on the initial droplet volume and substrate inclination. Kim et al.[38] explored experimentally the dynamics of isothermal droplet evaporation on a tilted substrate with a 90° droplet contact angle, and showed that the evaporation time of an inclined droplet increases as the gravitational influence due to inclination becomes stronger. Edwards et al.[42] observed the flow pattern and velocity inside tilted evaporating binary liquid droplets and concluded that both solutal Marangoni and buoyancy effects contributed to convection within the droplet.

Combining insights from these previous studies, two questions arise: 1) What are the evaporation dynamics on a heated vertical surface, and 2) can we use surface patterning to control or vary the evaporation process of evaporating droplets? In this paper we report an experimental study on water droplet evaporation on heated horizontal and vertical bi-phobic surfaces, consisting of a superhydrophobic matrix with a circular hydrophobic pattern and strong contact line pinning. We measured the droplet-substrate interfacial spatial temperature and heat flux distributions using infrared (IR) imaging for a full evaporation event. Analysis of synchronized optical images elucidates the coupling of fluid dynamics, *i.e.*, circulating convective currents within the droplet, and its thermal signature. We show that the droplets maintain high contact angles (characteristic of superhydrophobic surfaces) throughout most of the evaporation process, while remaining pinned (typical of smooth hydrophilic or rough surfaces with droplets in the Wenzel state). Even on vertical substrates, contact line pinning provides a high enough energy barrier for the droplet to overcome the increased gravitational influence and remain in a CCR evaporation mode. Interestingly, altered flow fields within the droplet and the gas phase decrease the total evaporation time for vertical droplets. Our understanding of droplet evaporation dynamics on patterned bi-phobic surfaces has the potential to influence and control evaporation dynamics in spray cooling and ink-jet printing applications.

■ **EXPERIMENTAL METHODS**

**Heater Design and Surface Processing.** The sample design, the fabrication process, and the corresponding wettability patterns are shown in Figure 1. The substrate consists of four layers. First, a 1.18 mm thick calcium fluoride window (UQG Optics) was coated with a 12 μm thin layer of black paint (Testors 18PK Black Enamel, 1149TT), which served as transducer for thermal imaging. The coating was applied *via* spin coating at 1200 rpm for 90 s and the thickness of the black paint layer was measured using a profilometer (Alpha-Step D-100 Stylus Profiler). On top of the black paint, a thin layer of chromium with 350 nm thickness was deposited using physical vapor deposition (PVD), serving as heater. Two triangular-cut copper sheets, attached to a DC power supply (Instek PSW 160-7.2), were bonded to the calcium fluoride window and in direct contact with the Cr layer to ensure a homogeneous temperature distribution in the center of the sample. The heated Cr surface measured 15.4 mm × 25.5 mm. The top side of the heater assembly was spray-coated using the commercially available superhydrophobic Glaco Mirror Coat.[25] At room temperature, the static contact angle of water droplets on Glaco is 164° with negligible contact angle hysteresis. Samples were allowed to dry for a minimum of 24 hours before further processing. To change the wettability of the surface, ultraviolet exposure (PSD Pro Series Digital UV Ozone System) was carried out for 105 minutes. An aluminum mask with a central hole of 1.5 mm diameter was used to cover the superhydrophobic surface. The wettability of the exposed area gradually decreased over time, reaching a sessile contact angle of 90-110° with strong contact line pinning (receding contact angle of 0°) after 105 minutes. The contact angle in the covered areas remained unchanged. After the exposure and removal of the mask, we obtained a bi-phobic surface consisting of a superhydrophobic matrix with a central circular hydrophobic spot (diameter: $D$ = 1.5 mm). The resulting apparent advancing and receding contact angles between the hydrophobic spot and the superhydrophobic background were 169° and 0°, respectively. Table S1 of the Supporting Information (SI) shows the decrease in contact angle with UV exposure time. It becomes clear that the droplets transition from a non-wetting Cassie state on the pristine Glaco coating to a wetting Wenzel state after UV exposure. Droplet contact angles were obtained from optical images using Image J ($\pm$ 3°).[43,44]

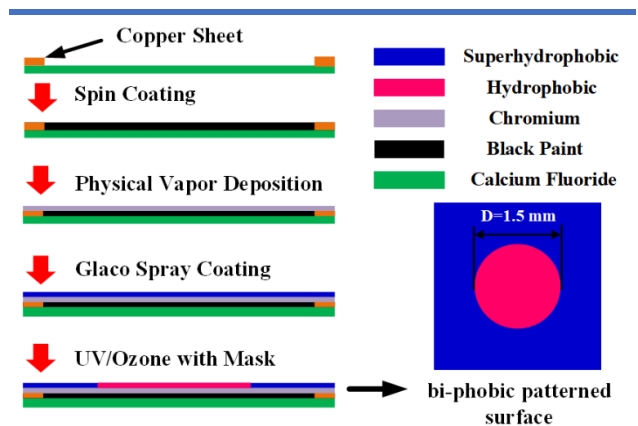

**Figure 1.** The process of surface wettability patterning.

**Experimental Setup.** Sessile droplets of de-ionized (DI) water, initially at room temperature and with a volume of 8 μl, were placed on heated bi-phobic substrates with an initial surface temperature of 74°C $\pm$ 1°C and allowed to evaporate. We conducted additional experiments at 45°C and 60°C to confirm that the observed evaporation dynamics are not unique to the chosen temperature level. However, we limit the presentation and discussion of results to surface temperatures of 74°C, as the IR camera signal-to-noise ratio is highest and evaporation processes and corresponding thermal-fluidic signatures are most pronounced at this temperature. Similar to Gibbons *et al.*[25] we did not observe a degradation of the surface wettability within the timeframe of these experiments. However, additional experiments at higher temperatures revealed a change in wettability, limiting our observations to the presented temperature. Experiments were conducted at least three times for each orientation and showed good repeatability.

A schematic of the experimental setup is depicted in Figure 2. Side-view images of the droplets were recorded using a Canon EOS Rebel T6i with a Canon MP-E 65mm f/2.8 1-5X Macro Lens at a spatial resolution of 2.3 μm/pixel. A white LED (Metabright Area Backlights) provided sufficient back-lighting for shadowgraph imaging. A Telops FAST M3$k$ infrared camera equipped with a 1x long working-distance lens (Telops) captured the temperature distribution of the sample in bottom-view using a gold-coated mirror (Thorlabs) at 1 frame/s with a resolution of 30 μm/pixel. For experiments with a vertical substrate orientation the mirror was removed and the IR camera pointed directly at the backside of the sample.

**Thermal Calibration and Heat Transfer Analysis.** To account for the deviation from black body emission of the black paint, for transmission losses through the calcium fluoride window, reflection losses from the mirror, and a thermal resistance from the Glaco coating, we calibrated the IR camera by attaching a surface RTD (Omega, PT100) to the top side of the sample. Through comparison with the IR camera readout we created a correction correlation, which relates the two temperature readings and is applied to each image before further processing.

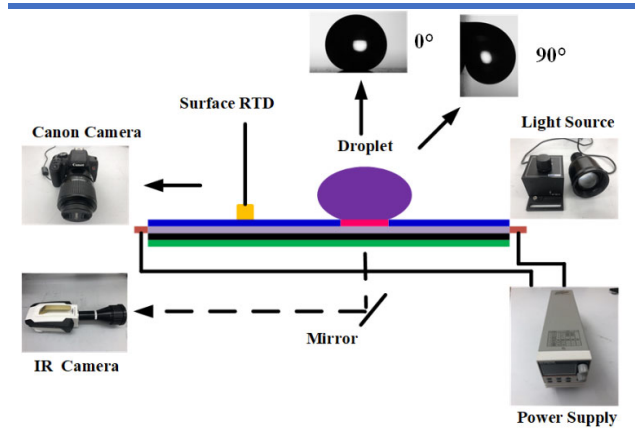

**Figure 2.** Schematic of the experimental setup.

In order to obtain heat transfer rates from the temperature distribution of the substrate, a transient energy balance was applied to each pixel element.[45–47] Detailed information on the analysis is presented in Section 2 of the SI. Briefly, the convective heat flux of each pixel element is expressed as $q''_{Top} = q_J/A - Q_{Store}/(\Delta t \cdot B^2) - q''_{Side} - q''_{Bottom}$, where $q''_{Top}$ is the heat flux to the top surface (i.e., droplet or air) of each pixel element, $q_J$ is the heat generated in the Cr thin film heater by Joule heating, $A$ is the total area of the heater, $Q_{Store}$ is the thermal energy stored in the pixel element in the time interval $\Delta t$ between two consecutive frames, $B$ is the size of a pixel element, $q''_{Side}$ is the heat flow due to conduction to/from the four neighboring pixel elements, and $q''_{Bottom}$ is the heat loss to the environment through conduction and natural convection at the bottom of the calcium fluoride window. The recorded thermal images were processed using a custom written MATLAB code to obtain temporal heat flux distributions. To account for slight differences in the fabrication process, the heat generation term was determined for each sample individually from an energy balance prior to droplet deposition, including natural convection and radiation heat transfer from the front side (assumption: emissivity Glaco $\varepsilon = 1$) and all other terms discussed above. Along with the measured electrical current, we find a total heater resistance of 8-9 Ω for all samples, in line with resistivity data for sputtered or e-beam deposited thin film Cr.[48] The resulting heat generation flux from Joule heating is approximately 0.93 kW/m². From a combination of the accuracy of the IR camera (30 mK), the use of a Gaussian filter (see SI), and the energy balance to determine the electrical resistance, the uncertainty in reported average heat flux values is 6.4 - 12%. Note that the error associated with the electrical resistance (6 - 8.4%) is the same for all data points on one sample and corresponds to an off-set in the heat transfer data.

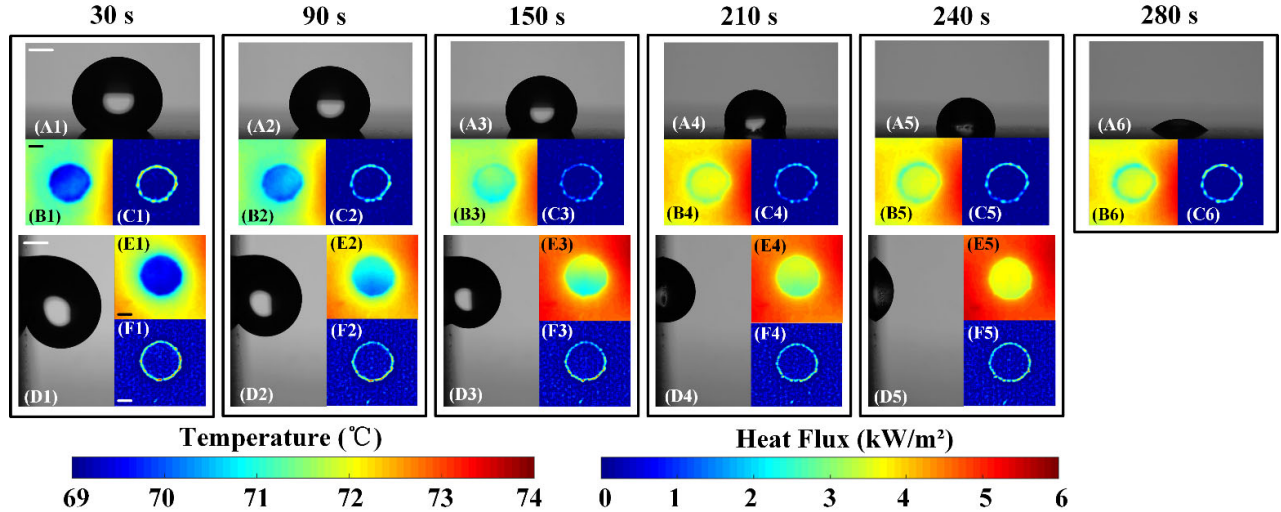

**Figure 3.** Optical images (A, D) and corresponding temperature (B, E) and heat flux (C, F) distributions of the evolution of sessile water droplets evaporating on heated horizontal (top; 0°) and vertical (bottom; 90°) bi-phobic surfaces. The total evaporation time is 293 s for the horizontal droplet and 265 s for the vertical droplet. Scale bars are 0.6 mm.

■ RESULTS AND DISCUSSION

**Qualitative observations.** Images from the simultaneous optical visualization of the droplet evaporation process in side-view, along with the corresponding temperature profiles of the substrate in bottom-view and the calculated local heat flux distributions are shown in Figure 3 for both horizontally and vertically mounted substrates. From these images, we can identify three main qualitative observations of the evaporation process: 1) Due to pinning of the contact line at the wettability boundary, contact angles decrease over time, while the wetted area remains approximately constant. Whereas contact angles are circumferentially symmetric for a droplet evaporating on a horizontal surface, contact angles are higher in the lower region of the droplet for vertical samples. At later stages of the evaporation process, contact angles at the upper and lower areas converge. 2) Similar to superhydrophobic surfaces,[21,25] the maximum local heat flux occurs in the three-phase contact line region during the entire evaporation process, despite the drastic change in contact angle over the course of the evaporation cycle. Josyula et al.[49] observed a strong correlation between the temperature difference within the droplet and the evolution of the droplet shape for the different modes of evaporation (*i.e.*, CCR, CCA, or stick-slip). A similar phenomenon is observed in our experiment. 3) The most important finding is the discrepancy in the thermal signature between the horizontal and vertical bi-phobic surfaces. On the horizontal surface, the droplet shape always resembles that of a spherical cap, and temperature and heat flux distributions are approximately symmetric and uniform. On the other hand, on the vertical bi-phobic surface, the temperature and heat flux distributions are not uniform in the intermediate stages of the evaporation process. We clearly see in panels E2/F2 through E4/F4 of Figure 3 that the lower part of the droplet is at a lower temperature and has a higher heat flux compared to the upper contact line region. Furthermore, droplets of the same volume evaporate faster on the vertical than the horizontal surface (265 s *vs.* 293 s). Please note that the initial fast increase in temperature for both substrate orientations is due to sensible heating as the droplet (initially at room temperature) is brought into contact with the hot surface. At later times the importance of sensible heating diminishes. As will be discussed below, sensible heating likely also plays a major role in establishing the initial non-uniform temperature distribution on the vertical surface. This spatially and temporally non-uniform thermal signature indicates that substrate orientation has a significant influence on the evaporation process of water droplets, in line with previous observations on the importance of droplet shapes on evaporation dynamics.[50,51] In the remainder of this paper we will discuss the influence of the orientation on contact angles, fluid motion, temperature and heat flux distributions, and their thermal-fluidic interplay in more detail.

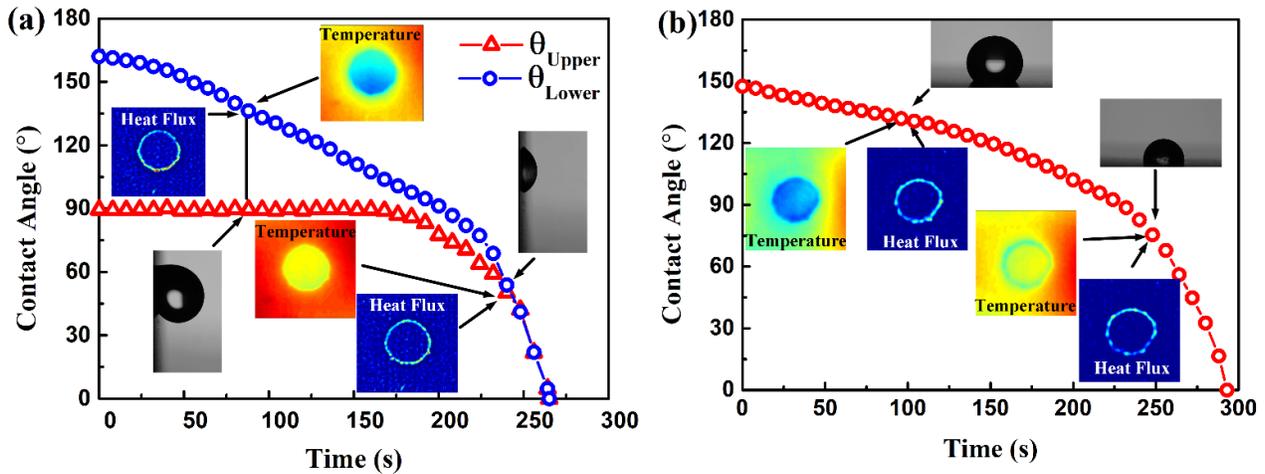

**Figure 4.** Evolution of the contact angle over time for (a) vertical and (b) horizontal bi-phobic surfaces.

**Analysis of droplet geometry and fluid motion.** Figures 4 and 5 show the variation of droplet contact angle and shapes with evaporation time on the vertical (a) and horizontal (b) bi-phobic surfaces. The insets in Figure 4 show the respective instantaneous optical images of the droplets, as well as temperature and heat flux distributions. As expected,[34] on the horizontal bi-phobic surface the contact angle decreases continuously as the droplet evaporates. Due to pinning of the contact line at the wettability boundary, the wetted radius remains approximately constant ($r = 0.75$ mm) throughout the entire evaporation process, and the droplet shape resembles that of a spherical cap. On the horizontal surface, droplet evaporation occurs in a clear CCR mode for entire duration of the evaporation event. In contrast to the CCA mode of a regular superhydrophobic surface, where the contact line of a droplet keeps receding, the contact radius here remains constant because of the pinning force at the wettability boundary, which obstructs the movement of the contact line. This variation in contact angle and evaporation mode likely influences the evaporation rate on the bi-phobic surface.[16,49,52]

The contact angle evolution on the vertical surface, however, exhibits three distinct stages, as can be seen in Figure 4a: (i) Initially, the contact angles at the upper and lower contact lines are $\theta_U \approx 90°$ and $\theta_L \approx 165°$, respectively. The weight of the water deforms the droplet to deviate from a spherical cap shape, as seen in Figure 5a. The balance of surface tension and gravity result in an initial wetted radius of 0.85 mm; larger than the size of the hydrophobic spot, with a bottom contact angle equal to the advancing contact angle on the pristine superhydrophobic surface. The upper contact line pinning provides a high enough energy barrier to prevent droplet sliding. During the first stage of the evaporation process ($t \leq 160$ s), the upper contact angle remains approximately constant, while the contact angle at the lower contact line decreases monotonically. At the same time, the bottom contact line recedes until being pinned at the wettability transition. (ii) During the second stage (160 s $< t \leq 240$ s), upper and lower contact angles decrease at a similar rate. (iii) Finally, for $t > 240$ s, the upper and lower contact angles become equal ($\theta \approx 53°$) and decrease sharply until the droplet evaporates completely. It is interesting to note that the temperature and heat flux profiles become circumferentially symmetric at this instance. The symmetry in the droplet geometry leads to a symmetry in the local thermal resistance and fluid flow fields, resulting in the symmetrical thermal signature. Based on the droplet profiles at early and intermediate times, we can divide the droplet into an upper region which has approximately 1/3 of the droplet volume, and a lower region which comprises roughly 2/3 of the droplet mass. Assuming the same heat flux through the substrate-droplet interface (based on Joule heating), this difference in thermal mass is likely responsible for the initial development of the vertically inhomogeneous temperature distribution observed in Figure 3, panels E2-4. A more detailed discussion on the interplay between contact angles, temperatures, and heat flux can be found in Section 4 of the SI.

To quantify the effect of droplet shape on convective currents, we seeded evaporating droplets with natural graphite powder (10 μm spherical particles). The experimental setup was similar to that shown Figure 2, but without IR camera and mirror. Videos were taken with a Canon EOS Rebel T6i with a Canon MP-E 65mm f/2.8 1-5X Macro Lens at 30 fps. At a 3x magnification and an aperture of f/3.2 the depth of field (DoF) is approximately 0.1 mm. The local velocity field was measured by tracing particles within the focal plane, which was aimed at the center of the droplet (highest contrast, *i.e.* sharp edges, at the droplet periphery). Streamlines were recorded experimentally by tracking individual particles over an extended period of time using ImageJ, as shown in Figure 5c,d. Using Snell-Descartes' law of refraction,[53] we estimate the error of reported particle velocities to approximately 12%. For the vertical bi-phobic surface, one small-size oval vortex is observed in the upper region of the droplet, with high fluid velocities in the upper region ($v > 2.2$ m/s) and weaker flow in the lower region ($v < 0.75$ m/s). In contrast, the flow pattern on the horizontal bi-phobic surface is Y-shaped and bifurcated, with two symmetric vortices on both sides of the droplet.[54] Supplementary videos are available to visualize these flow fields.

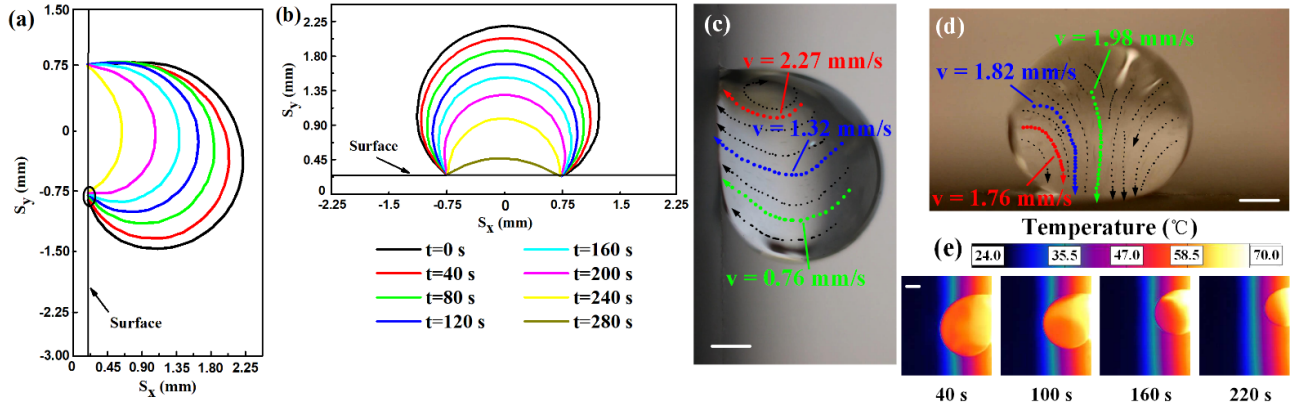

**Figure 5.** (a,b) The time-varying outline of droplets and (c,d) image composition of the overlay of recorded particle trajectories including selected velocities for droplets evaporating on vertical and horizontal bi-phobic surfaces. (e) Side-view temperature profiles of a droplet on the vertical surface. Note that the temperature color scale is from raw IR camera data and has not been calibrated. For a more detailed temporal evolution of the side-view temperature distribution, see Movie 3 of the SI. Scale bars are 0.5 mm.

Three distinct physical phenomena contribute to the observed convective currents: Capillary flow, Marangoni flow, and thermal buoyancy flow. While the strength of capillary flow is related to the evaporative flux, the relative strength of Marangoni and thermal buoyancy flows can be analyzed by comparing the Marangoni number *Ma* and the Rayleigh number *Ra*.[49,55] Based on the observed velocities, we get $Ra = 3.3 \times 10^4$ and $Ma = 2.4 \times 10^5$. The ratio $Ra/Ma = 0.135 \ll 1$ suggests that buoyancy is negligible compared to surface-tension driven Marangoni flow. However, the theoretically calculated Marangoni number often over-predicts the Marangoni strength in experiments for water droplets due to surface contamination.[56] We hence conclude that both thermal buoyancy and thermal Marangoni flow contribute to the observed convective current, especially for the horizontal sample where the temperature gradient across the droplet (solid-liquid interface to bulk liquid) is parallel to gravity. For droplets on the vertical bi-phobic surface, the temperature gradient across the droplet is nearly perpendicular to gravity, meaning that buoyant convection likely becomes negligible.

Instead, for the vertical bi-phobic surface, the small-size oval vortex near the upper contact line is formed from the co-action of Marangoni and capillary flow. At the droplet surface, due to a negative correlation between surface tension and temperature, Marangoni forces result in flow from the low-surface region (*i.e.*, contact line) to the high-surface tension region (*i.e.*; apex).[57] Inside the droplet, capillary flow establishes to replenish evaporating liquid near the contact line. As discussed earlier, the lower thermal mass in the upper region of the droplet results in higher relative temperatures, which are also evident in the side-view IR images in Figure 5e. This temperature field and the evaporative flux create a self-sustained oval vortex in the upper region. Near the lower contact line region, we only observe capillary flow towards the contact line. Due to a higher thermal mass and an increased conduction resistance across the droplet, temperatures remain relatively low in this region. Low-temperature liquid from the bulk region is continuously pushed towards the solid-liquid interface, maintaining the vertical temperature profile in the droplet, and suppressing Marangoni flow in the lower region.

**Analysis of temperature and heat flux distributions.** After characterizing the droplet geometry and fluid motion, we now turn to quantifying their influence on the thermal signatures of the evaporating droplets. Figure 6a shows the time evolution of the temperature and heat flux of the upper and lower contact line regions on a vertical bi-phobic surface, respectively. In contrast to the lower contact line region, where temperature and heat flux vary linearly with time (after the initial sharp increase in temperature when the droplet is being brought into contact with the substrate), the changes of temperature and heat flux in the upper contact line region show two different trends. During the first period ($t$ < 100 - 120 s), the temperature of the upper contact line region rises more quickly than that of the lower contact line region, followed by a relatively long period of modest increase. We attribute this initial strong increase in temperature to the smaller thermal mass of the upper part of the droplet and the associated lower energy required for sensible heating. At later times, conduction, convection, and evaporating in the upper region stabilize the temperature profile, leading to a stagnation or even slight increase of the heat flux. At $t \approx 125$ s the heat flux is $0.98 \pm 0.11$ kW/m² and then gradually increases to $1.08 \pm 0.13$ kW/m² towards the end of the evaporation process. On the one hand, due to a monotonic increase in droplet temperature (*i.e.*, sensible heating), we would expect the heat transfer between the heated substrate and the droplet to decrease monotonically over time as the thermal driving potential decreases, as observed for the lower contact line area. On the other hand, an elevated droplet temperature enhances the evaporative flux (*i.e.*, latent cooling).[17] We thus observe a competition between sensible heating and latent cooling.

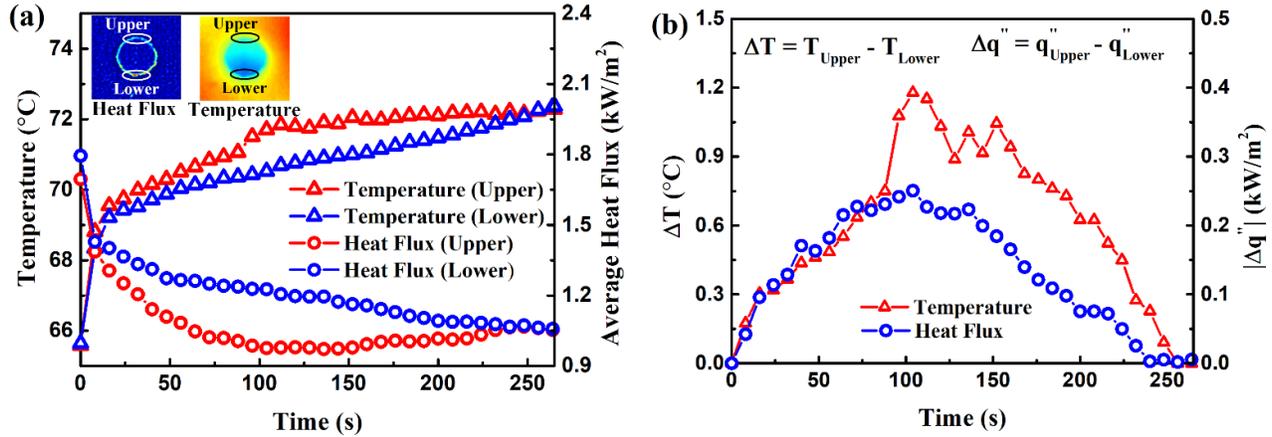

Figure 6. Time evolution of the average temperature and heat flux on a vertical bi-phobic surface for (a) the upper and lower contact line regions, respectively, and (b) the absolute difference between the two regions.

Counterintuitively, the average heat flux is higher at the lower contact line than at the upper contact line, as can be seen in Figure 6. Xu *et al.*[27] found that the diffusion of vapor away from droplets on superhydrophobic surfaces is significantly restricted by a high contact angle. For moderate contact angles (*i.e.*, $\theta < 90°$), vapor diffusion from the liquid-air interface is geometrically not restricted, whereas the diffusion resistance becomes important at higher contact angles, thus limiting the evaporative mass flux. Similarly, at higher contact angles, the thermal conduction path through the droplet to the liquid-vapor interface increases.[58] In both cases, we would expect a decrease in heat flux with increasing contact angle at the bottom of the droplet. However, our experiments show that the heat flux in the upper region, which has the smaller contact angle, is smaller than that of the lower contact line region with the larger contact angle. This is in contrast to the current understanding on the influence of contact angles on evaporation rates. We propose two possible reasons for this apparent contradiction. First, as discussed above, the non-symmetric convective flow pattern within droplet on the vertical substrate replenishes the lower part of the droplet with colder liquid from the bulk *via* capillary flow, leading to stronger perpendicular temperature gradients and hence enhanced convective and conductive heat transfer. Second, and more importantly, buoyancy effects in the surrounding gas phase and the temperature distribution at the droplet surface contribute to the higher-than-expected evaporative flux at the lower contact line. In the surrounding gas, buoyancy-driven flow from the lower region towards the upper region, caused by heated gas near the substrate and from the density difference between moist and dry air, contributes to a larger concentration gradient at the lower contact line, which can overcome the geometric diffusion limit. In order to quantify the influence of buoyancy flow in the vapor domain on evaporation rates, we conducted additional numerical simulations using COMSOL based on a quasi-steady state assumption. The evaporation is predicted by incorporating the thermal diffusion model with convection in the vapor domain.[59,60] The simulation methodology and a detailed discussion of the results are provided in Section 5 of the SI. The numerical results show a strong upward buoyancy flow (~0.1 m/s) around the droplet. Effected by the flow field, the vapor concentration distribution surrounding the droplet becomes "flame-shaped", with a large vapor-saturated region above the droplet and a compressed region below the droplet. This leads to a weak vapor concentration gradient at the upper part and an enhanced gradient in the wedge beneath the droplet. Meanwhile, the lower droplet temperature at the bottom leads to a smaller saturation concentration, enhancing the sharper vapor gradient. Thus, the diffusion resistance near the upper contact line region is increased, whereas the diffusion resistance near the lower contact line region is lowered and the evaporation rate enhanced for the same contact angle as compared to a horizontal surface. In the symmetry plane of the droplet, the predicted concentration gradient at the lower contact line is 42.3% larger than at the upper contact line, which leads to a 30.9% higher evaporative flux. Once the droplet profile becomes symmetric ($t \approx 240$ s), the heat flux difference between upper and lower contact line vanishes, consistent with our hypothesis. As contact angles converge, the flow pattern becomes more symmetric, assimilating temperature and heat transfer rates in the top and bottom regions. In the meantime, the decreased liquid-vapor interfacial area results in an overall smaller evaporation rate, which reduces the vapor concentration around the droplet and decreases the strength of buoyant flow in the gas.

Finally, we compare the average temperature and total heat flux for the entire droplet for the two substrate orientations. Typical applications of droplet evaporation include high-power cooling devices, and as such a lower average temperature and higher overall heat flux are desirable. Figure 7a shows the magnitude of the average temperature and total heat flux to the evaporating droplet on the vertical bi-phobic surface. As the droplet heats up due to sensible heating and a decrease in volume, the driving force for heat transfer, *i.e.*, the temperature difference between droplet and substrate decreases monotonically, and so does the heat flux. Contrary to that, the average temperature and heat flux on a horizontal surface shows three distinct phases, as shown in Figure 7b. After the initial intense sensible heating of the droplet (phase I), the evaporation process equilibrates and the heat flux remains approximately constant (phase II). As the contact angle decreases, the thinning of the droplet (less volume but same contact area) leads to an increase in heat transfer (phase III). Despite the increase in heat flux towards the end of the evaporation process on the horizontal surface, the total heat transfer on the vertical surface is always larger than that on the horizontal surface, leading to shorter droplet evaporation times on vertical bi-phobic surfaces (compare to Figure 3).

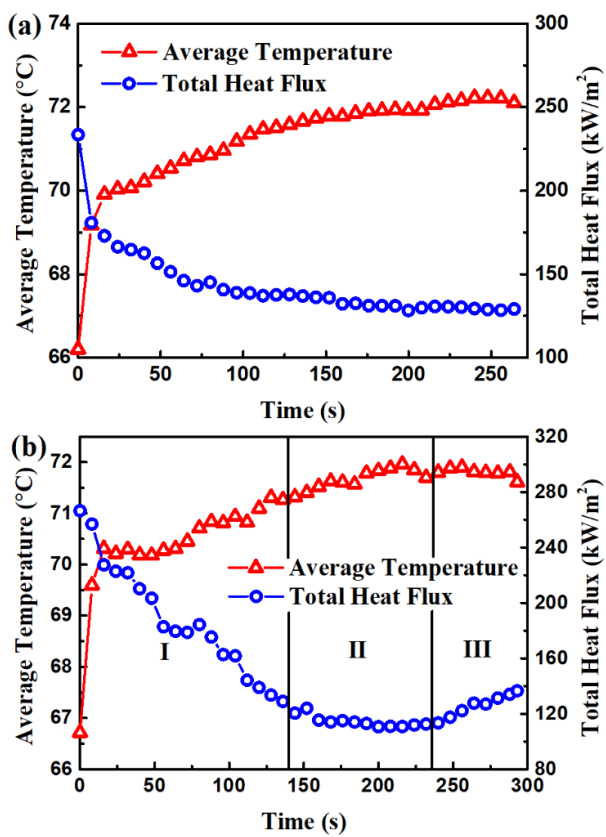

**Figure 7.** Evolution of the average temperature and total heat flux for droplets on (a) vertical and (b) horizontal bi-phobic surfaces.

### CONCLUSIONS

To summarize, we studied the evaporation of water droplets on heated bi-phobic surfaces (superhydrophobic matrix with circular hydrophobic patterns). In particular, we focused on the influence of substrate orientation (horizontal or vertical) on the evaporation dynamics. We found that heat transfer is more efficient on vertical bi-phobic surfaces due to altered convective flow fields within the droplet and in the surrounding gas, leading to higher overall heat transfer rates and shorter evaporation times. While horizontally placed droplets maintain a symmetrical shape and uniform temperature and heat transfer distribution during evaporation, droplets on a vertical substrate are asymmetric and thermally inhomogeneous. Our main findings are threefold: (i) Droplets always evaporate in the CCR mode on bi-phobic surfaces. Attributed to the pinning force at the wettability boundary, the contact radius remains fixed, leading to the highest heat transfer rates at the three-phase contact line region for both vertical and horizontal surfaces. (ii) Compared to a droplet evaporating on a horizontal surface, the temperature and heat flux distributions on the vertical surface are not spatially uniform in the intermediate stages of the evaporation process. Counterintuitively, we observe a lower temperature and higher heat flux in the lower region, despite a higher contact angle. We propose that buoyancy effects and the temperature distribution at the droplet surface result in a higher concentration gradient beneath the droplet, enhancing evaporation at the lower contact line. Temperature and heat flux become circumferentially uniform once the contact angles in both regions become equal. (iii) The total thermal energy transfer to the droplet is more efficient (up to 10% shorter evaporation times) on a vertical substrate compared to that on a horizontal substrate. While the increased wetting radius near the lower contact line on the vertical surface can contribute to this increase in heat transfer, we mainly attribute this efficiency increase – normalized by the initial droplet volume – to the altered temperature and heat flux distributions and the competition of convection, conduction, and evaporation for the non-symmetrical vertical droplet.

### ■ ASSOCIATED CONTENT

**Supporting Information**
Supporting Information (PDF)
    Section 1: The effect of UV exposure on contact angles
    Section 2: Heat flux calculation
    Section 3: Property value
    Section 4: Contact angle - heat flux dependence
    Section 5: Simulation methodology and results
    Section 6: Nomenclature
Movie S1: Particle tracer movement for an evaporating droplet on a horizontal bi-phobic surface. (AVI)
Movie S2: Particle tracer movement for an evaporating droplet on a vertical bi-phobic surface. (AVI)
Movie S3: Side-view surface temperature evolution on a vertical surface. (AVI)

### ■ AUTHOR INFORMATION

**Corresponding Author**
*E-mail: p.weisensee@wustl.edu. Tel: +1-314-935-7951.

**Notes**
The authors declare no competing financial interest

### ■ ACKNOWLEDGMENTS

The authors would like to thank Gongyu Tang and Jianxing Sun for helping with the construction of the setup and valuable discussions. The authors acknowledge financial support from Washington University in St. Louis and the Institute of Materials Science and Engineering for the use of instruments and staff assistance. Wenliang Qi acknowledges financial support from China Scholarship Council (CSC).

# Supporting Information to:

# Evaporation of sessile water droplets on horizontal and vertical bi-phobic patterned surfaces

Wenliang Qi,[†,‡] Junhui Li,[†] Patricia B. Weisensee[†,§,*]

[†]Department of Mechanical Engineering & Materials Science, Washington University in St. Louis, MO, USA
[‡]College of Power and Energy Engineering, Harbin Engineering University, Harbin, China
[§]Institute of Materials Science and Engineering, Washington University in St. Louis, MO, USA

* Corresponding author: p.weisensee@wustl.edu

## 1. The effect of UV exposure on contact angles

**Table S1**. Effect of UV exposure on contact angles. Please note that static and advancing contact angles have a high uncertainty (~10-15°) as a function of exposure time, as derived from 3 independent measurements. Receding contact angles converge to 0° very rapidly and are consistently observed for 45+ minutes of exposure. For the experiments discussed in the main manuscript, this receding contact angle plays the important role of pinning the contact line at the wettability boundary.

| UV Exposure Time (min) | Sessile static contact angle (°) | Advancing contact angle (°) | Receding contact angle (°) |
|---|---|---|---|
| 0 | 164 | 166 | 158 |
| 30 | 148 | 155 | 58 |
| 60 | 132 | 148 | 0 |
| 90 | 119 | 137 | 0 |
| 105 | 112 | 122 | 0 |
| 130 | 107 | 119 | 0 |
| 105 (1.5mm spot) | - | 169[a] | 0 |

[a] The maximum contact angle for the droplet to pin on the spot

## 2. Heat Flux Calculation

In order to obtain the local heat flux distribution from the heater to the fluid, an unsteady energy balance was applied at each pixel element. The pixel element was defined as the portion of the heater that is mapped to one pixel in the image of the IR camera with the thickness of the chromium layer and black paint. Thus, the pixel element is a cuboid, as illustrated in Figure S1.

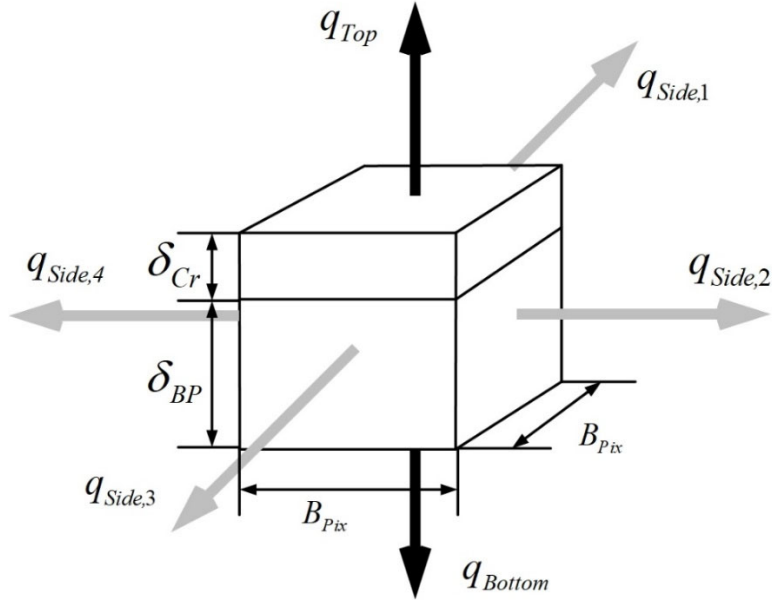

*Figure S1: Contributing heat flows for a single pixel element*

Energy conservation equations applied to each pixel element can be expressed as:
$$Q_{Top} = Q_J - Q_{Store} - Q_{Side} - Q_{Bottom}. \qquad (1)$$
The calculation of the heat flux is based upon several assumptions and simplifications:
- Due to the small thickness of the black paint and chromium layers, the thermal resistance is negligible and their thermal properties are bulked together into one representative variable.
- The heat generation from Joule heating is homogenous in the entire chromium layer.
- The temperature of the heater (including the chromium and the black paint) is uniform across the heater thickness and within each pixel element.

These assumptions and simplifications result in the following expression for the heat flux at the top of the pixel element (*i.e.*, heat flux to air or the droplet):
$$q''_{Top} = \frac{q_J}{A_s} - \frac{Q_{Store}}{B_{Pix}^2 \cdot \Delta t} - q''_{Side} - q''_{Bottom}. \qquad (2)$$

All variables are defined in section 6: *Nomenclature* of this document.

### 2.1. Joule Heating

The chromium and black paint serve as a resistive heater. The heat generated in the Cr thin film by Joule heating can be expressed as:
$$q_J = I_s^2 R_{Cr}. \qquad (3)$$

### 2.2. Stored Heat

The heat stored in the pixel element was calculated using the temperature change of the element with time:
$$Q_{Store} = B_{Pix}^2 \left( \delta_{Cr} \rho_{Cr} c_{p,Cr} + \delta_{BP} \rho_{BP} c_{p,BP} \right) \left( T_{x,y,\tau+1} - T_{x,y,\tau} \right). \qquad (4)$$

In order to minimize the signal noise, a weighed averaging of the pixel temperature in time

was performed:
$$T_{x,y,\tau} = 0.25T_{x,y,\tau-1} + 0.5T_{x,y,\tau} + 0.25T_{x,y,\tau+1}. \tag{5}$$

### 2.3. Lateral Conduction

Heat is conducted to the neighboring pixel elements at the four sides of each element. The application of the 2D heat diffusion equation leads to

$$q''_{Side} = -(k_{Cr}\delta_{Cr} + k_{Bp}\delta_{Bp})\left(\frac{\partial^2 T_s}{\partial x^2} + \frac{\partial^2 T_s}{\partial y^2}\right), \tag{6}$$

with

$$\frac{\partial^2 T_s}{\partial x^2} \approx \frac{T_{x+1,y,\tau} - T_{x,y,\tau}}{\Delta x^2} + \frac{T_{x-1,y,\tau} - T_{x,y,\tau}}{\Delta x^2}, \tag{7}$$

$$\frac{\partial^2 T_s}{\partial y^2} \approx \frac{T_{x,y+1,\tau} - T_{x,y,\tau}}{\Delta y^2} + \frac{T_{x,y-1,\tau} - T_{x,y,\tau}}{\Delta y^2}. \tag{8}$$

As in this method only five different pixel elements (the one of interest and 4 neighboring) are used for the calculation of the surface heat flux, it is very prone to noise, as the heat conduction term is extremely sensitive to the spatial signal noise of the input temperature field. Similar to ref. [1], we thus applied a Gaussian filter to the temperature signal which can suppress the sensitivity. The size of the filter was $n = 5$ and the standard deviation was $\sigma_n = 2.5$.

### 2.4. Bottom Losses

Based on the thermal resistance network, the heat flux through the bottom of the heater is given by:

$$q''_{Bottom} = \left(\frac{1}{h_{air}} + \frac{\delta_{CF}}{k_{CF}}\right)^{-1}(T_s - T_{air}). \tag{9}$$

The natural convection heat transfer coefficient from the bottom of the heater assembly can be estimated by using McAdams's correlation[2] for natural convection from a downward-facing horizontal plate:

$$h_{air} = \frac{Nu \cdot k_{air}}{L}, \tag{10}$$

$$Nu = 0.27 \cdot Ra_L^{1/4}, \tag{11}$$

$$Ra = \frac{g\beta(T_s - T_{air})L^3}{\gamma_{air}\alpha_{air}}, \tag{12}$$

$$L \equiv \frac{A_s}{P}. \tag{13}$$

Due to the small magnitude of natural convection from the back side of the substrate, the same set of equations is used for a vertical plate without introducing significant uncertainty.

## 3. Property Values

The chromium, black paint[3], calcium fluoride and air properties are outlined in Table S2. Experimental parameters are listed in Table S3.

**Table S2.** Thermo-physical properties.

| | | |
|---|---|---|
| Chromium density, $\rho_{Cr}$ | 7190 | $kg/m^3$ |
| Chromium thermal conductivity, $k_{Cr}$ | 90.7 | $W/(m \cdot K)$ |
| Chromium specific heat, $c_{Cr}$ | 448 | $J/(kg \cdot K)$ |
| Black paint density, $\rho_{BP}$ | 1261 | $kg/m^3$ |
| Black paint thermal conductivity, $k_{Bp}$ | 0.095 | $W/(m \cdot K)$ |
| Black paint specific heat, $c_{BP}$ | 2835 | $J/(kg \cdot K)$ |
| Calcium fluoride thermal conductivity, $k_{CF}$ | 9.71 | $W/(m \cdot K)$ |
| Air thermal conductivity, $k_{air}$ | $3.005 \times 10^{-2}$ | $W/(m \cdot K)$ |
| Air kinematic viscosity, $\gamma_{air}$ | $2.055 \times 10^{-5}$ | $m^2/s$ |
| Air thermal diffusivity, $\alpha_{air}$ | $2.94 \times 10^{-5}$ | $m^2/s$ |

**Table S3.** Experiment parameters.

| | | |
|---|---|---|
| Calcium fluoride thickness | 1.18 | mm |
| Black paint thickness | 12 | μm |
| Chromium layer thickness | 350 | nm |
| Heated surface area | 15.4 × 25.5 | mm² |
| IR camera resolution | 320 × 256 | pixel (30 μm/pixel) |
| IR camera framerate | 1 | fps |
| High-speed camera resolution | 1920 × 1080 | pixel (2.3 μm/pixel) |
| High-speed camera framerate | 30 | fps |
| Water droplet volume | 8 | μl |

## 4. Contact angle - heat flux dependence

To further elucidate the influence of contact angles on the thermal-fluidic behavior of the evaporating droplet, we plot temperature and heat flux as a function of contact angle for the upper (Figure S2a) and lower (Figure S2b) contact line regions on a vertical bi-phobic surface separately. In the upper part of the droplet, temperature and heat flux remain nearly constant for all contact angles (except for the rapid initial temperature increase due to sensible heating, which is not plotted here for clarity). This is surprising, given that previous studies observed an increase in evaporation flux with decreasing contact angles.[4,5] We attribute this to the non-homogeneous concentration gradient in the gas phase, which is the rate-limiting factor.

In contrast, for the lower contact line region, we observe a strong dependence of temperature and heat flux on contact angles for $\theta_L > 90°$, as shown in Figure S2b. We identify three phases: In phase I, the droplet heats up rapidly from its initial temperature (≈ 25°C) to slightly below the substrate temperature due to sensible heating while the heat flux decreases. With decreasing contact angle (*i.e.*, as time evolves), the heat flux continues to decrease monotonically during phase II, until both temperature and heat flux become constant in phase III for $\theta_L < 90°$. This also marks the point when the droplet shape resumes to be symmetric (compare to Figures 3, 4 and 5 of the main manuscript). Based on a vapor-diffusion model, Dash and Garimella[6,7] found that a contact angle of 90° marks a turning point in the evaporation dynamics, where the evaporative flux is highest at the apex for $\theta > 90°$, and highest near the contact line for $\theta < 90°$. We also observe distinct temperature and heat flux distributions for contact angles larger and smaller than 90°, respectively. However, as discussed in the main manuscript, for evaporation on vertical substrates we observe a contrary effect to what is expected: decreasing or constant heat fluxes with decreasing contact angles ($\theta > 90°$ and $\theta < 90°$, respectively), which ties back to the observation made above for the upper contact line.

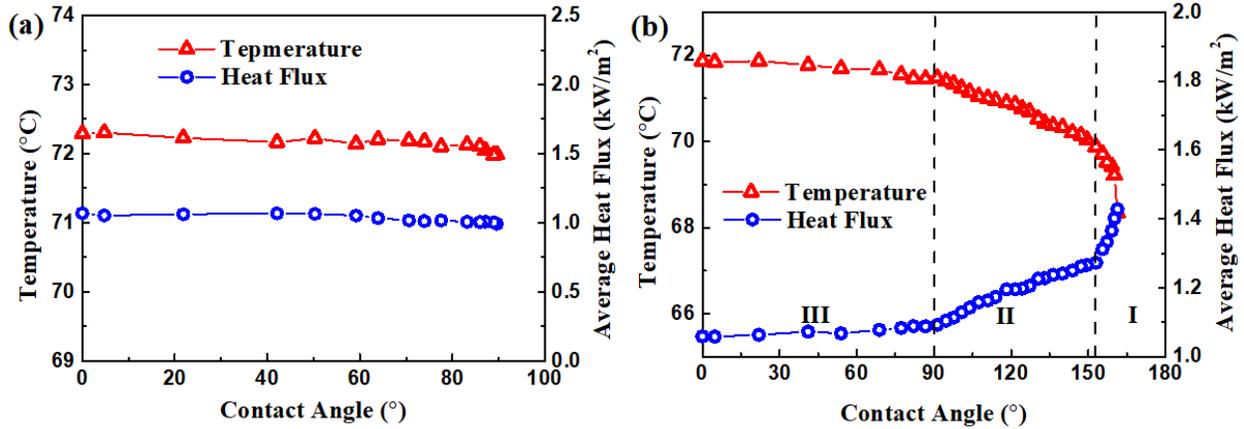

*Figure S2: Evolution of temperature and heat flux as a function of contact angle on the vertical bi-phobic surface for the (a) upper (initial rapid increase in temperature due to sensible heating not shown) and (b) lower contact line. Note that the time evolution of the evaporation process is right to left (i.e., contact angles decrease over time).*

## 5. Simulation methodology and results

The goal of the numerical simulation is to help explain the non-uniform heat flux distribution observed in the experiment of a vertical substrate, and not to reproduce the entire evaporation process. The numerical model used is the same as in the authors' previous work[8,9]. Because the inner convective current is well-analyzed experimentally in the main manuscript, here we neglect the convection flow inside the droplet and assume a quasi-steady diffusion and heat transfer process at an evaporation time $t = 40s$. The thermal buoyancy flow in the vapor domain is calculated to demonstrate its influence on the non-uniform heat flux, especially near the upper and lower contact line regions.

### 5.1. Mathematical model and boundary conditions

When analyzing the mass transport in the vapor domain, the process is governed by the gas species transport equation, which is given by

$$-\vec{V} \cdot \nabla C_v + \nabla \cdot (D \nabla C_v) = 0. \tag{14}$$

A constant concentration boundary condition was assigned at the far field, where the concentration was equal to the concentration of vapor in air at 25% relative humidity and 22°C. An impenetrable condition (*i.e.*, zero diffusive flux) was assigned to the solid-vapor interfaces, which can be expressed as

$$\frac{dC_v}{d\vec{n}} = 0. \tag{15}$$

Because of the large droplet size (8 µL), the liquid-vapor interfacial pressure was not affected by the curvature of the droplet. Therefore, the concentration at the liquid-vapor interface was assumed to be equivalent to the saturation concentration, *i.e.*,

$$C_{v,lv} = C_{sat}(T_{lv}). \tag{16}$$

Consequently, the heat transfer in the liquid is governed by the heat conduction equation (neglecting thermal convection, as mentioned above):

$$\nabla^2 T = 0. \tag{17}$$

In the simulation, a temperature boundary condition, derived from the experimentally determined temperature distribution described in the main manuscript, was assigned to the solid-liquid interface at the hydrophobic spot. The temperature boundary condition for the substrate and droplet-substrate interface is taken from the IR thermography images, with $T_{sub} = 72°C$ and $T_{drop-sub} = 69.5°C$ at evaporation time $t = 40s$, as shown in Figure S3. At the liquid-vapor interface, a heat flux thermal boundary condition was imposed to satisfy the energy conservation between conduction and evaporation heat transfer:

$$k_l \left(\frac{\partial T}{\partial \vec{n}}\right)_{lv} = h_{fg}[-D\left(\frac{\partial C_v}{\partial \vec{n}}\right)_{lv} + \vec{V}_n C_v]. \tag{18}$$

Considering thermal buoyancy and Stefan flow, the convective flow is included in the gas domain by incorporating the continuity, momentum, and energy equations in the simulation:

$$\nabla \cdot (\rho \vec{V}) = 0, \tag{19}$$

$$\rho(\vec{V} \cdot \nabla)\vec{V} = \nabla \cdot \left[-p\mathbf{I} + \mu\left[\nabla\vec{V} + (\nabla\vec{V})^T\right] - \frac{2}{3}\mu(\nabla \cdot \vec{V})\mathbf{I}\right] + \rho\mathbf{g}, \tag{20}$$

$$\rho c_p \vec{V} \cdot \nabla T - \nabla \cdot (k\nabla T) = \rho c_p \frac{\partial T}{\partial t}. \tag{21}$$

The liquid-vapor interfacial temperature is solved based on conduction through the droplet and is used as the boundary condition for solving the temperature in the gas domain. The velocity boundary condition at the liquid-vapor interface in the gas domain is given by[10]

$$\vec{V}_n = \frac{1}{C_{air}} \cdot D\frac{\partial C_{air}}{\partial n} = -\frac{1}{C_a - C_v} \cdot D\frac{\partial C_v}{\partial n}. \tag{22}$$

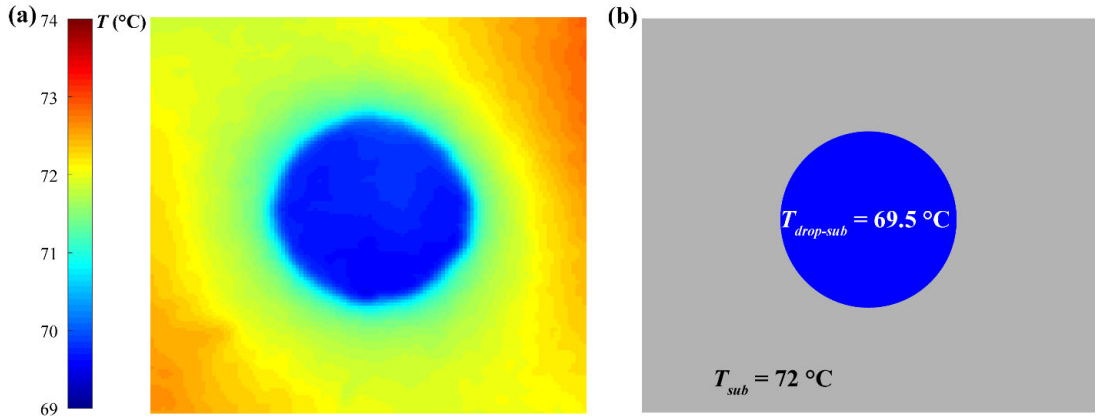

*Figure S3: The temperature at the substrate and the droplet-substrate interface during droplet evaporation on a vertical bi-phobic surface at evaporation time t = 40s. (a) IR thermography image. (b) Boundary conditions assigned for the simulation.*

### 5.2. Geometry and computational domain

The equilibrated droplet shape on the vertical surface is a deformed non-spherical shape due to gravitational distortion. The experimental droplet shape at evaporation time $t = 40$ is shown in Figure S4a, and the droplet geometry calculated by Surface Evolver[11] is shown in Figure S4b. The generated droplet geometry is then imported into COMSOL, where a hemispherical far-field (over 30 times larger than the droplet size) is assigned as system boundary.

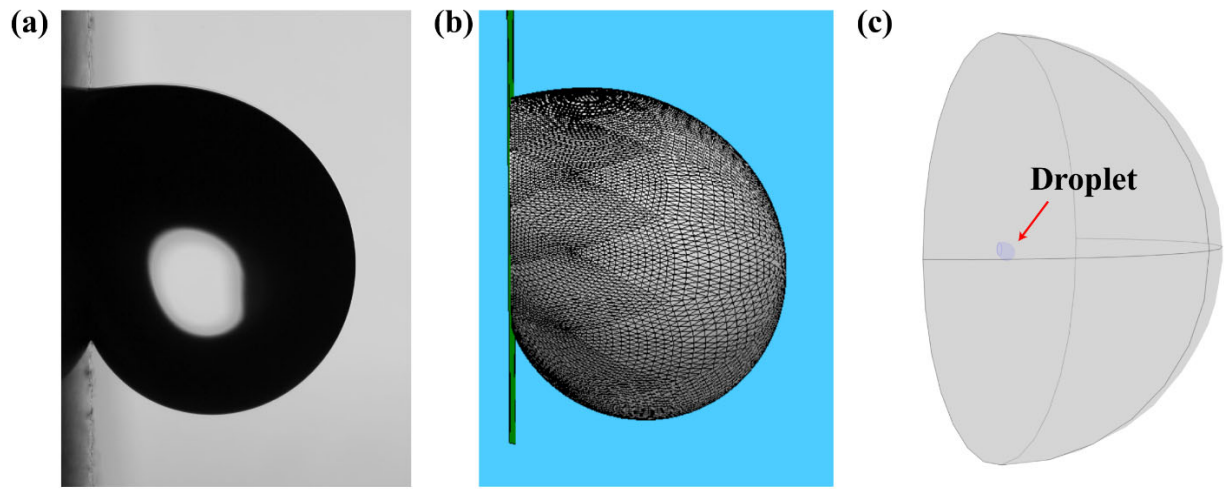

*Figure S4: Geometry and computational domain for the simulation of droplet evaporation on a vertical bi-phobic surface at evaporation time t = 40s. (a) Experimental droplet shape. (b) Calculated droplet shape using Surface Evolver. (c) Computational domain to numerically solve for droplet evaporation.*

## 5.3. Simulation results

The vapor concentration at the droplet symmetry plane is shown in Figure S5a. The vapor concentration distribution is strongly influenced by gas flow, as shown in Figure S5b. This buoyancy flow consists of thermal buoyancy due to gas heating near the substrate and due to the lower density of moist air with respect to dry air. This convection results a higher vapor concentration region above the droplet, whereas the concentration in the lower region of the droplet is reduced. In addition to the buoyancy flow, the droplet surface near the lower contact line has a lower temperature due to the larger thermal mass ($\rho V c_p$) of the bulk region of the droplet. The lower liquid-vapor interfacial temperature results in a lower saturated vapor concentration at the droplet surface, which provides a larger concentration gradient at the lower contact line than at the upper contact line (Figure S5a). In the symmetry plane, the concentration gradients at the upper and lower contact lines are predicted to be $6.91 \times 10^5 \; mol/m^4$ and $9.84 \times 10^5 \; mol/m^4$, respectively. While the favorable contact angle at the upper contact line promotes high evaporation rates there as expected, the strong concentration gradient at the lower contact line region enhances evaporation strongly as compared to evaporation on a horizontal surface with the same contact angle. The total evaporation fluxes (diffusive and convective flux) at the upper and lower contact line are $2.14 \; mol/(m^2 \cdot s)$ and $2.80 \; mol/(m^2 \cdot s)$ respectively. Although we neglect convection within the droplet, these numerical results match the experimental observations well, which show that the lower contact line region has a higher evaporative flux and hence higher heat transfer through the substrate-droplet interface.

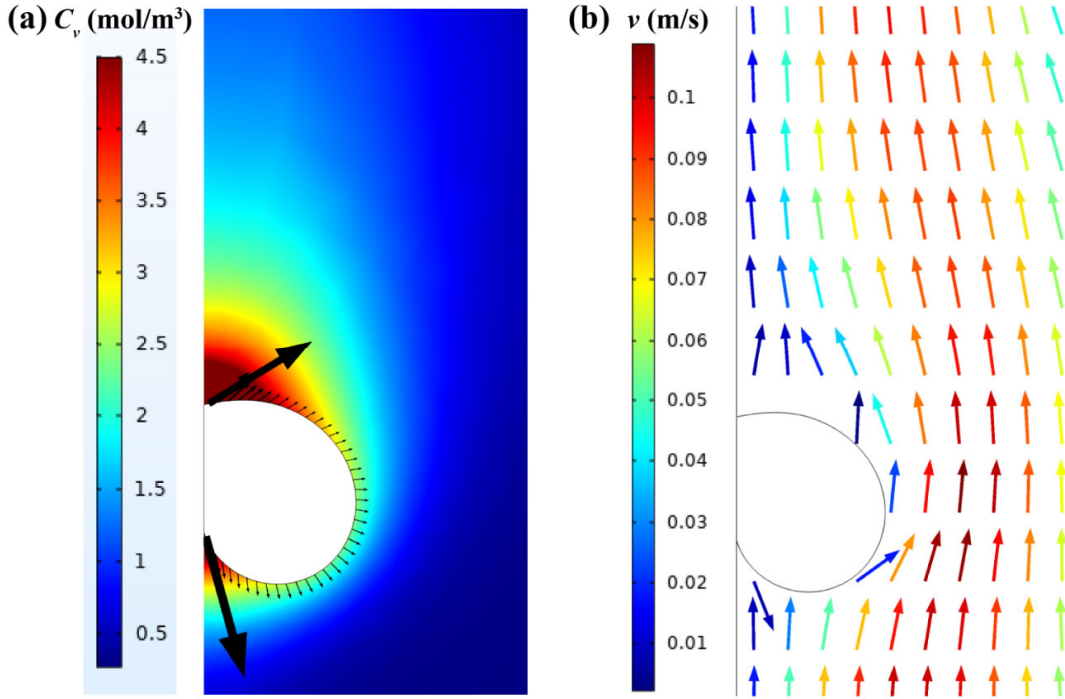

*Figure S5: Simulation results of droplet evaporation on a vertical bi-phobic surface at evaporation time t = 40s. (a) Contour of vapor concentration (filled color) and concentration gradient (black arrow) of the symmetry plane. The size of the arrows is proportional to the concentration gradient magnitude; the arrow direction is reversed. (b) Vector plot of the gas flow field around the droplet.*

## 6. Nomenclature

| | | | |
|---|---|---|---|
| $A$ | Area $[m^2]$ | $\alpha$ | Thermal diffusivity $[m^2/s]$ |
| $B$ | Pixel length $[m]$ | $\beta$ | Thermal expansion coefficient $[K^{-1}]$ |
| $c_p$ | Specific heat capacity $[J/(kg \cdot K)]$ | $\delta$ | Thickness $[m]$ |
| $C$ | Concentration $[mol/m^3]$ | $\gamma$ | Kinematic viscosity $[m^2/s]$ |
| $D$ | Diffusion coefficient $[m^2/s]$ | $\mu$ | Dynamic viscosity $[Pa \cdot s]$ |
| $g$ | Gravitational constant $[m/s^2]$ | $\rho$ | Density $[kg/m^3]$ |
| $h$ | Heat transfer coefficient $[W/(m^2 K)]$ | $\tau$ | Time step of camera (1/framerate) |
| $h_{fg}$ | Heat of vaporization $[W/kg]$ | _Subscripts_ | |
| $I$ | Electrical current $[A]$ | $air$ | Air |
| $k$ | Thermal conductivity $[W/(m \cdot K)]$ | $BP$ | Black paint |
| $L$ | Characteristics length $[m]$ | $Bottom$ | Bottom surface of heater |
| $P$ | Perimeter $[m]$ | $CF$ | Calcium fluoride |
| $q$ | Rate of heat transfer $[W]$ | $Cr$ | Chromium |
| $q''$ | Heat flux $[W/m^2]$ | $drop$ | Droplet |
| $Q$ | Thermal energy $[J]$ | $g$ | Gas (Air and Vapor) |
| $R$ | Chromium resistance $[\Omega]$ | $J$ | Joule heating |
| $t$ | Time $[s]$ | $lv$ | Liquid-vapor interface |
| $T$ | Temperature $[K]$ | $Pix$ | Pixel |
| $v$ | Velocity $[m/s]$ | $s$ | Heater surface |
| $\vec{V}$ | Velocity vector | $sub$ | Substrate |
| $V$ | Voltage $[V]$ or Volume $[m^3]$ | $Side$ | Side surface of heater |
| _Dimensionless numbers_ | | $Store$ | Capacitance or storage |
| $Nu$ | Nusselt number | $Top$ | Top surface of heater |
| $Ra$ | Rayleigh number | $v$ | Vapor |
| _Greek symbols_ | | $x,y$ | Coordinates |